\documentclass[aps,prd,twocolumn,showkeys,floats,showpacs, nofootinbib]{revtex4-1}
\usepackage{graphics,epsfig,color} 
\usepackage{amsmath,amssymb}

\usepackage{enumerate}
\usepackage{aas_macros}
\usepackage{bibentry}
\usepackage[colorlinks=true]{hyperref}

\def\L {\mathcal{L}}
\def\O { \mathcal{O}}
\def\s{\star}
\def\c{\diamond}
\def\infeq {\underset{t \rightarrow \infty}{=}}

\newcommand{\p}[1]{\dot{#1}}
\newcommand{\pp}[1]{\ddot{#1}}

\def\be{\begin{equation}}
\def\ee{\end{equation}}
\def\bea{\begin{eqnarray}}
\def\eea{\end{eqnarray}}
\begin{document}
\title{Late-time cosmology of scalar-tensor theories with universal multiplicative coupling between the scalar field and the matter Lagrangian}
\author{Olivier Minazzoli}
\email{ominazzoli@gmail.com}
\affiliation{UMR ARTEMIS, CNRS, University of Nice Sophia-Antipolis,
Observatoire de la C\^ote d'Azur, BP4229, 06304, Nice Cedex 4, France}

\author{Aur\'elien Hees}
\email{aurelien.hees@gmail.com}
\affiliation{Department of Mathematics, Rhodes University, Grahamstown 6140, South Africa}

\begin{abstract}
We investigate the late-time cosmological behaviour of scalar-tensor theories with a universal multiplicative coupling between the scalar field and the matter Lagrangian in the matter era. This class of theory encompasses the case of the massless string dilaton (see Damour and Polyakov, General Relativity and Gravitation, 26, 1171) as well as a theory with an intrinsic decoupling mechanism in the solar system (see Minazzoli and Hees, Phys. Rev. D 88, 041504). The cosmological evolution is studied in the General Relativity limit justified by solar system constraints on the gravitation theory. The behaviour of these cosmological evolutions are then compared to two types of observations: the constraints on temporal variations of the constants of Nature and the distance-luminosity measurements. In particular, the non-minimal coupling implies that the distance-luminosity relation is modified compared to General Relativity. Theories producing a cosmological behaviour in agreement with these observations are identified.
\end{abstract}
\pacs{}
\maketitle

\section{Introduction}
Today, gravitation is facing a major problem: on one hand, general relativity (GR) has passed all the stringent solar system experiments \cite{Will_book93,*Will-lrr-2006-3} ; on the other hand, GR and the standard model of particles are not sufficient to explain certain galactic or cosmological observations. The most widespread solution consists of extending the matter-energy content of the Universe by introducing Dark Matter and Dark Energy. Another possibility consists of modifying the law of gravitation at large scales without introducing any new type of matter/energy. The number of alternative theories of gravity developed in the last years has been growing very fast (for a wide review, see \cite{clifton:2012fk}). 

Amongst all the alternative theories of gravity, the most widespread are scalar-tensor theories of gravitation. Although scalar-tensor theories are often considered with a minimal scalar-to-matter coupling, scalar-tensor theories with non-mininal coupling generically appear in (gravitational) Kaluza-Klein theories with compactified dimensions \cite{uzanLRR11,fujiiBOOKst,*overduinPRep97} and in string theories at the low energy limit \cite{uzanLRR11,DamPolyGRG94,*DamPolyNPB94,Sstring88,beanPLB01,gasperiniPRD02,damourPRD02,*damourPRL02}; but also in $f(R)$ gravity~\cite{nojiriPLB04,*allemandiPRD05,*nojiriARXIV06,*bertolamiPRD07,* bertolamiPRD08,* bertolamiCQG08, *sotiriouCQG08, *de-felice:2010uq, *sotiriouRMP10,*nojiriPR11,*harkoPRD13,*taminiPRD13}, in Brans-Dicke-like theories \cite{dasPRD08,*bisabrPRD12,*moffatIJMPD12}, in massive theories of gravity \cite{dvali:2000dq,*nicolis:2009cr}, or in the so-called MOG (MOdified Gravity) \cite{moffatJCAP06,*moffatCQG09}. Besides, it has aslo been argued that requirering gauge and diffeomorphism invariances would single out such types of theories as well \cite{armendarizPRD02}. 

Moreover, cosmological observations of Dark Energy are quite often explained by a scalar field \cite{ratra:1988vn,*caldwell:1998kx,*peeblesRMP03,elizaldePRD08}, and the inflation paradigm also introduces such a field \cite{guth:1981uq,*linde:1982kx,*albrecht:1982vn,*linde:2008ys,*bruneton:2012kx,elizaldePRD08}. Finally, variations of the constants of Nature (such as the fine structure constant \cite{damourNPB96,fujiLNP04,*petrovPRC06,*guena:2012hl,*webbPRL01,*murphyMNRAS03,*marionPRL03,*bizePRL03,*fischerPRL04,*peikPRL04,*olivePRD04,*webbPRL99,*srianandPRL04,*chandAA05} for example) are usually modeled with a scalar field as well~\cite{bekenstein:1982zr,*sandvik:2002ly,*dvaliPhRL02,*oliverPRD08,uzanLRR11,damour:2012zr,barrow:2013}.

However, the introduction of the scalar field has to satisfy stringent solar system constraints on gravity \cite{Will_book93,*Will-lrr-2006-3,adelberger:2009fk,*wagnerCQG12,*adelbergerARNP03,*adelbererPRL07,*hees:2012fk,*hees:2014yu,williams:2009ys}. In this context, several mechanisms have been proposed to screen the scalar field or to naturally decouple it from matter. For instance, screening mechanisms (chameleons \cite{khoury:2004uq,*khouryPRD04,*hees:2012kx}, symetron \cite{hinterbichler:2010fk,*hinterbichler:2011uq} or Vainshtein mechanism \cite{vainshtein:1972ve,*deffayet:2002ly}) are different ways to reduce the effects of the scalar field in some regions of space~\cite{khoury:2010zr}. Recently, we propose a new decoupling mechanism of the scalar field in region of space-time where the pressure is negligible (such as in in the solar system or during the late-time cosmology) \cite{nousPRD13}. For this reason, we propose to dubb such a scalar-field \textit{pressuron}. In \cite{nousPRD13}, we saw that the pressuron theory naturally passes all solar system tests on the post-Newtonian phenomenology. 

But in addition to satisfying the solar system tests of gravitation, the developed theory has to explain the late-time cosmological observations. In particular, it is known that the Universe is currently experiencing an acceleration of the cosmic expansion which has been inferred from  distance-luminosity versus redshift measurements done with Supernovae Ia  (SNe Ia) \cite{riess:1998ly,*perlmutter:1999ve}. It is interesting to study if the developed scalar-tensor theories of gravitation are able to reproduce such an acceleration of the cosmic expansion without the introduction of a cosmological constant. On the other hand, scalar-tensor theories induce a variation of various fundamental constant of nature such as the gravitational constant $G$, but even sometimes the fine structure constant $\alpha$, or the weak interaction constant $\alpha_W$. But the spatial and temporal variations of these constants are severely constrained by observations and experiments (see \cite{williams:2009ys,kaspiAJ94,*guentherAJ98,*copiPRL04} for G, \cite{damourNPB96,fujiLNP04,*petrovPRC06,*guena:2012hl,webbPRL01,*murphyMNRAS03,marionPRL03,*bizePRL03,*fischerPRL04,*peikPRL04,*olivePRD04,*webbPRL99,*srianandPRL04,*chandAA05} for $\alpha$, or \cite{damourNPB96,malaneyRP93,*reevesRMP94} for $\alpha_W$). Therefore, all scalar-tensor theories have to converge during the evolution of the Universe toward non-variation of the fundamental constants in order to satisfy the constraints coming from observations. A converging mechanism towards GR has been found for usual scalar-tensor theories \cite{damourPRL93,*damourPRD93,jarvPRD08,jarvPRD10,jarvPRD12}, as well as dilaton-like theories \cite{DamPolyGRG94}, but is effective for some specific classes of theories only depending on the specific coupling function of the theory \cite{gerard:1995fk,*gerard:1997xy,*oukouiss:1997kq,sernaCQG02}.

In this communication, we focus on the study of the late-time cosmology of scalar-tensor theories of gravity with a universal multiplicative coupling between the scalar field and the matter Lagrangian. This encompasses the case of the massless dilaton considered in \cite{DamPolyGRG94,*DamPolyNPB94} and the case of the pressuron \cite{nousPRD13}. In order to do so, we apply the method recently developed in \cite{jarvPRD12} for the case of usual scalar-tensor theories in the general relativity limit. The idea is to study the possible cosmological evolution of the scalar field and of the cosmic scale factor by solving perturbatively the field equation (a different approach consisting in using a phase space approach is used in \cite{barrow:2013}). Then, we can identify the conditions under which the scalar field evolves towards an attractor, which is required in order to satisfy constraints on variations of fundamental constants. Then, we would like to see if the same kind of cosmological evolution is able to reproduce SNe Ia data. Since the coupling between the scalar field and matter is non-minimal, the expressions of the observables are not necessarily the same as in GR. Therefore, we derive the expression of the distance-luminosity versus redshift from first principles. We show that the scalar field explicitly enters the expression of the distance-luminosity. Therefore, SNe Ia data might be explained by the modification of the distance-luminosity relation instead of being explained by an acceleration of the cosmic scale factor. 

This paper derives several important results. First, we show that a general multiplicative scalar-to-matter coupling (such as for the massless dilaton defined in \cite{DamPolyGRG94,*DamPolyNPB94}) leads to the same cosmological behaviour as in scalar-tensor theories with minimal scalar-to-matter coupling, but with different parameters. This means that the evolution depends on the choice of the scalar coupling function and can lead to a convergence towards GR. Second, we show that the pressuron dynamic freezes for any scalar-field coupling function enabling a natural explanation of the apparent present constancy of fundamental constants. Unfortunately, the modification of the distance-luminosity relation does not allow to explain SNe Ia observations by itself, with a scalar field converging towards a constant. Therefore, a potential is still needed to explain the apparent acceleration of the cosmic expansion with the pressuron.

In section \ref{sec:eq} we present the general action considered in this paper and we derive the cosmological field equations. In section \ref{sec:evol}, we solve the field equations using a perturbative method similar to the one used in \cite{jarvPRD12}. A non-perturbative approach based on \cite{sernaCQG02} for the case without potential is also presented. In section \ref{sec:obs}, we compute the observables from first principles and compare their evolutions with observations. Finally, we conclude in section \ref{sec:concl}.

\section{Cosmological equations}
\label{sec:eq}
Let us consider the action of a class of scalar-tensor theories with a universal coupling between the scalar field and the material Lagrangian:
\begin{eqnarray}
S&=& \frac{1}{c}\int  d^4x \sqrt{-g} \Big[ f(\Phi) \mathcal{L}_m (g_{\mu \nu}, \Psi) + \label{eq:actiondila}  \\ 
&&\quad \frac{1}{2\kappa}\left(\Phi  R-\frac{\omega(\Phi)}{\Phi} (\partial_\sigma \Phi)^2-V(\Phi) \right)\Big] . \nonumber
\end{eqnarray}
where $R$ is the Ricci scalar constructed from the metric $g_{\mu \nu}$, $g$ is the metric determinant, $\kappa=\frac{8\pi G}{c^4}$, with $G$ the gravitational constant \footnote{Note however that it is different from the effective constant measured with Cavendish-type experiments. See sec. \ref{sec:effect_grav_const} for a discussion on the effective constant of gravitation.} and $c$ the velocity of light in vacuum, $V(\Phi)$ is the scalar-field potential, $f(\Phi)$ is an arbitrary adimensional function, $\mathcal{L}_m$ is the matter Lagrangian and $\Psi$ represents the non-gravitational fields. It has to be noted that such an action encompasses the effective string theory's low energy action at tree level, but also the assumed full loop expansion considered as a toy model in \cite{DamPolyGRG94} (for which $f(\Phi) \propto \Phi$ and $V(\Phi)=0$, see Appendix \ref{app:dilaton}). The action (\ref{eq:actiondila}) covers also the theory studied in \cite{nousPRD13} for which $f(\Phi) \propto \sqrt{\Phi}$ and $V(\Phi)=0$. As shown in \cite{nousPRD13}, there is a decoupling of the scalar field in region where the pressure is negligible in this specific theory. Therefore, we dubb this particular scalar field \emph{pressuron}. In particular, this theory naturally satisfies solar system tests of gravitation \cite{nousPRD13}. 

The definition of the stress-energy tensor is given by
\begin{equation}
T_{\mu \nu}=-\frac{2}{\sqrt{-g}} \frac{\delta(\sqrt{-g}\mathcal{L}_m)}{\delta g^{\mu \nu}}.
\end{equation}
From the extremization of the action (\ref{eq:actiondila}), one gets the following Einstein field equations
\begin{eqnarray}
R_{\mu \nu}-\frac{1}{2}g_{\mu \nu}R= \kappa~ \frac{f(\Phi)}{\Phi}T_{\mu \nu}+ \frac{1}{\Phi} [\nabla_{\mu} \nabla_{\nu} -g_{\mu \nu}\Box]\Phi  \nonumber\\
+\frac{\omega(\Phi)}{\Phi^2}\left[\partial_{\mu} \Phi \partial_{\nu} \Phi - \frac{1}{2}g_{\mu \nu}(\partial_{\alpha}\Phi)^2\right]-g_{\mu \nu} \frac{V(\Phi)}{2 \Phi} , \label {eq:motiong}
\end{eqnarray}
and the Klein-Gordon equation for the scalar field
\bea
\label{eq:motionPhi}
\frac{2\omega(\Phi)+3}{\Phi}\Box \Phi= \kappa \left( \frac{f(\Phi)}{\Phi} T - 2 f'(\Phi) \mathcal{L}_m \right) \\ - \frac{\omega'(\Phi)}{\Phi} (\partial_\sigma \Phi)^2 + V'(\Phi) - 2 \frac{V(\Phi)}{\Phi}  \nonumber
\eea
where $T$ is the trace of the stress-energy tensor and the prime denotes the derivation with respect to the scalar field.

The invariance of action (\ref{eq:actiondila}) under diffeomorphisms imply the following conservation equation :
\be
\nabla_\sigma T^{\mu \sigma} = \left(\L_m g^{\mu \sigma} - T^{\mu \sigma} \right) \partial_\sigma \ln{f} \label{eq:non_conserv}\, .
\ee
For a perfect fluid respecting the conservation equation, the stress-energy tensor writes  $T^{\alpha \beta}=(\epsilon+P)U^\alpha U^\beta + P g^{\alpha \beta}$ while the Lagrangian is  $\L_m = -\epsilon$ \cite{moiPRD12,moiPRD13}, where $\epsilon$ and $\rho$ are the total and rest mass energy densities and $U^\sigma$ is the four-velocity of the fluid. Let us emphasize that this Lagrangian is only valid for a perfect fluid. In particular, it is not valid for electromagnetic radiation which is characterized by $\mathcal L_{\textrm{EM}}=0$ in vaccuum. In this communication, we are interested in the late-time cosmological evolution and therefore, we will consider only a fluid of dust characterized by $P=0$. The radiation is negligible in this part of the cosmological evolution and will not be considered here.

Considering a flat Friedmann-Lema\^itre-Robertson-Walker (FLRW) metric for the Universe\footnote{In the following, we are using $c=1$.}
\be
ds^2=- dt^2+a^2(t) \left[dx^2+dy^2+dz^2 \right],
\ee
and a perfect fluid, the field equations (\ref{eq:motiong}) become
\begin{subequations}\label{eq:field1}
\bea
H^2=\kappa \frac{f(\Phi)}{3 \Phi} \epsilon+ \frac{\omega(\Phi)}{6} \left(\frac{\p{\Phi}}{\Phi} \right)^2-H \frac{\p{\Phi}}{\Phi} + \frac{V(\Phi)}{6\Phi}, \label{eq:comsoNE1}
\eea
\bea
2\p{H}+3H^2&=& -2 H \frac{\p{\Phi}}{\Phi}-\frac{\omega(\Phi)}{2} \left(\frac{\p{\Phi}}{\Phi} \right)^2- \frac{\pp{\Phi}}{\Phi} \nonumber \\
&&+\frac{V(\Phi)}{2\Phi}- \kappa\frac{f(\Phi)}{\Phi} P,\label{eq:comsoNE2}
\eea
where $H$ is the Hubble function defined as $H \equiv \p{a}/a$ and the dot denotes the derivative with respect to the cosmic time $t$.
The Klein-Gordon equation (\ref{eq:motionPhi}) for the scalar field reduces to
\bea
\pp{\Phi}&=&-3 \p{\Phi} H+ \frac{A(\Phi)}{2} \left(2 \omega(\Phi)+3 \right) {\p{\Phi}^2} \nonumber \\
& +& \frac{\kappa f(\Phi)}{2 \omega(\Phi)+3 } \left[\left(1 - 2 \frac{\Phi f'(\Phi)}{f(\Phi)} \right)\epsilon - 3  P\right]  \label{eq:comsoNE3}\\
&+& \frac{1}{2 \omega(\Phi)+3 } \left[2V(\Phi)-\Phi V'(\Phi) \right] \nonumber,
\eea
\end{subequations}
where $A(\Phi)$ is defined as in \cite{jarv:2010fk,jarvPRD12} by 
\be\label{eq:defA}
A(\Phi)= \frac{d}{d \Phi} \left(\frac{1}{2 \omega(\Phi)+3} \right)=-\frac{2\omega'(\Phi)}{(2\omega(\Phi)+3)^2}.
\ee
Meanwhile, the conservation equation (\ref{eq:non_conserv}) reduces to \footnote{It has to be noted that the simple (usual) form of the conservation equation arises in the present case from an exact cancellation in the development of equation (\ref{eq:non_conserv}) and therefore is rather remarkable.}:
\be
\p{\epsilon} + 3 \frac{\p{a}}{a} (\epsilon+P) =0.
\ee
We shall set $f(\Phi) \propto \Phi^n$, where $n \in \mathbb{R}$, such that it encompasses the cases considered both in \cite{DamPolyGRG94} and \cite{nousPRD13}. The assumptions allows to write
\be
 1 - 2 \frac{\Phi f'(\Phi)}{f(\Phi)} = 1-2n \, .
\ee
Following the development made in \cite{jarvPRD12}, for dust matter (ie. $P=0$), we use the redundancy of the system of Eqs.~(\ref{eq:field1}) in order to eliminate $\epsilon$. It leads to the following two equations
\begin{subequations}\label{eq:cosmo_decoup}
\bea
\pp{\Phi}&+&3H\p{\Phi}=\frac{A(\Phi)}{2}\left(2 \omega(\Phi)+3\right) {\p{\Phi}}^2 \label{eq:cosmoDecP} \\
&&+\frac{2V(\Phi)-\Phi V'(\Phi)}{2\omega(\Phi)+3} \nonumber\\
&&+ \frac{1-2n}{2 \omega(\Phi)+3} \left(3 \Phi H^2 + 3 H \p{\Phi} - \frac{\omega(\Phi)}{2}\frac{{\p{\Phi}}^2}{\Phi} -\frac{V(\Phi)}{2} \right),\nonumber
\eea
and
\bea
&&\p{H}=-\frac{3}{2} H^2+H \frac{\p{\Phi}}{2 \Phi}-\frac{\omega(\Phi) }{4}\frac{{\p{\Phi}}^2}{\Phi^2} \label{eq:cosmoDecH}\\
&&-\frac{1}{4}\left(2 \omega(\Phi)+3\right)A(\Phi) \frac{{\p{\Phi}}^2}{\Phi} +\frac{V(\Phi)}{4 \Phi}- \frac{V(\Phi)-\Phi V'(\Phi)/2}{2 \omega(\Phi)+3} \nonumber \\
&&-  \frac{1-2n}{2\left(2 \omega(\Phi)+3\right)} \left(3  H^2 + 3 H \frac{\p{\Phi}}{\Phi} - \frac{\omega(\Phi)}{2}\frac{{\p{\Phi}}^2}{\Phi^2} -\frac{V(\Phi)}{2\Phi} \right). \nonumber 
\eea
\end{subequations}
It has to be noted that these equations only slightly differ form the usual scalar-tensor case with minimal coupling ($f(\Phi)=1$) considered in \cite{jarvPRD12} (that is recovered for $n=0$ and $V(\Phi)=0$).

\section{Solution of the field equations}\label{sec:evol}
In this section, we will solve the field equations in order to derive the cosmological evolution of the scalar field and of the scale factor. This will allow us to determine under which conditions the considered theory converges towards GR. In a first step, we present an analytical perturbative approach following what is done in \cite{jarvPRD12}. This perturbative scheme can be used in the general case covered by the action (\ref{eq:actiondila}). We will use it to consider first a massless scalar field ($V(\Phi)=0$) and then extend the results in the case of a self interacting scalar field. However, we also present a non-perturbative analytical procedure in the case without potential. The method followed in this case is inspired by \cite{sernaCQG02}.

\subsection{Perturbative approach in the GR limit and the no potential case} \label{sec:pertGR}
In this section, we follow the approach presented in \cite{jarvPRD12} and we consider that there is no potential in the action (\ref{eq:actiondila}). In the following, we study the behaviour of the late-time cosmological evolution in the matter era in the GR limit. This limit is justified in particular by solar system tests of gravity. The GR limit is mathematically defined by 4 conditions (see \cite{jarv:2010fk,jarvPRD10,jarvPRD12}):
\begin{enumerate}[ (a):]
	\item $\frac{1}{2\omega(\Phi)+3}\rightarrow 0$ motivated by solar system tests of gravity (although they do not hold for the pressuron, see the discussion in Sec. \ref{sec:pressuron}). We define $\Phi_\s$ by
	\begin{equation}
		\frac{1}{2 \omega(\Phi_\star)+3}=0 \label{eq:condclose2GR}.
	\end{equation}
	\item $\dot\Phi \rightarrow 0$ motivated by the constraints on the variation of fundamental constants.
	\item $A_\star \equiv A(\Phi_\star)\neq 0$.
	\item $\frac{1}{2 \omega(\Phi)+3}$ is differentiable in $\Phi_\star$.
\end{enumerate}

It is worth mentioning that these conditions are consistent with the field equations as one can check with Eqs. (\ref{eq:field1}) (see also the discussion in \cite{jarvPRD12}). With these assumptions, we can develop our field equations around a background such that
\be \label{eq:expansion}
\Phi(t)=\Phi_\star+x(t), \qquad H(t)=H_\s(t)+h(t),
\ee
where $H_\s(t)$ is the Hubble function corresponding to the evolution in GR, while $x(t)$ and $h(t)$ are small perturbations. 

With the assumptions mentioned above, one has
\begin{subequations}
\be
\frac{1}{2 \omega(\Phi)+3}=\frac{1}{2 \omega(\Phi_\s)+3}+A_\s x + \mathcal{O}(2) =A_\s x + \mathcal{O}(2),
\ee
and
\be
(2 \omega(\Phi)+3) \p{\Phi}^2 = \frac{\p{x}^2}{A_\s x} + \O(2).
\ee
\end{subequations}
As in \cite{jarvPRD12}, the zeroth order solution of (\ref{eq:cosmoDecH}) gives 
\begin{equation}\label{eq:HS}
	H_\s(t) = \frac{2}{3t}=H_{\rm GR}(t)
\end{equation}	
where one has set the integration constant time equal to 0 for convenience and where $H_{\rm GR}$ stands for the classical GR evolution with no cosmological constant. The first order of Eqs. (\ref{eq:cosmo_decoup}) respectively write
\begin{subequations}\label{eq:pertGR}
\be
\pp{x}(t)+3 H_\s(t) \p{x}(t)= \frac{{\p{x}^2(t)}}{2 x(t)} + 3 (1-2n) A_\s \Phi_\s~  H^2_\s (t) x(t), \label{eq:cosmox}
\ee
and
\bea
\p{h}(t)+3 H_\s(t)h(t)= - \frac{1}{4 \Phi_\s} \left(1+\frac{1}{2 A_\s \Phi_\s} \right) \frac{{\p{x}^2(t)}}{x(t)}\label{eq:cosmoh} \nonumber\\ 
+ \frac{1}{2\Phi_\s} H_\s(t) \p{x}(t) - \frac{3}{2} (1-2n) A_\s H^2_\s(t) x(t)\, .
\eea
\end{subequations}
 For $n=0$, we recover the  results from \cite{jarvPRD12}. Let us nevertheless mention that an additional condition (not mentioned in \cite{jarvPRD12}) is necessary in order to get Eq. (\ref{eq:cosmox}): $\dot x \ll H_\star \Phi_\star$. This condition gives the limit at which the perturbative approach used here breaks down.

\subsubsection{Solutions of the perturbative equations}
The solutions of Eqs. (\ref{eq:pertGR}) have been derived and discussed in details in \cite{jarvPRD12} for $n=0$. The solutions in the general case $n\neq 0$ are similar except from the fact that the critical parameter $D$ is modified and now writes
\be \label{eq:D}
D=1+\frac{8}{3} (1-2n) A_\s \Phi_\s \, ,
\ee
instead of $D=1+8/3  ~A_\s \Phi_\s$ in the usual scalar-tensor case \cite{jarvPRD12}. In the following, we will briefly review the various possible cosmological evolutions but we refer to \cite{jarvPRD12} for a complete detailed discussion. 

For the sake of conciseness, the exact solutions of Eqs.~(\ref{eq:pertGR}) are given in Appendix~\ref{app:gen}. They depend on the critical parameter $D$ (\ref{eq:D}):
\begin{itemize}
\item $D> 0$: The solutions are polynomial (see Eqs. (\ref{eq:solDpos})).  Let us mention that the behaviour depends highly on the value of $D$. If $D>1$, the cosmological evolutions will diverge from the GR evolution and the approximation scheme used here will eventually break down. If $D<1$, the solutions will asymptotically converge towards GR: the scalar field will tend to a constant and the Hubble parameter tends towards its GR expression. The case $D=1$ is not allowed if $n\neq \frac{1}{2}$ since it contradicts the assumption (c). The case $n=\frac{1}{2}$ corresponding to the pressuron leads to $D=1$ independently of the function $\omega(\Phi)$ and is considered in details in Sec. \ref{sec:pressuron}.

\item $D= 0 :$ The solutions are logarithmic (see Eqs.~(\ref{eq:solD0})) and converge asymptotically towards GR.

\item $D<0 :$ The solutions are damped oscillations (see Eqs.~(\ref{eq:solDneg})). The behaviour of these solutions is developed into details in \cite{jarvPRD12}. Basically, they converge towards GR solution in the manner of damped oscillations.
\end{itemize}

In conclusion, when $D>1$ the solutions diverge from GR and when $D<1$ the solutions converge towards GR with different behaviour depending on the value of $D$.

\subsubsection{Discussion on the limitations of the perturbative approach}

As noticed in \cite{jarvPRD12}, in order to develop perturbatively the field equations, one needs to assume $A_\s \neq 0$ (see assumption (c) above). However, such a condition is quite limiting regarding the possible coupling function $\omega(\Phi)$ considered. Indeed, the condition writes
\be
A(\Phi_\s)= - \frac{\omega'(\Phi_\s)}{(2\omega(\Phi_\s)+3)^2} \neq 0 \label{eq:condDEV}
\ee
while $\Phi_\s$ satisfy the condition (\ref{eq:condclose2GR}). For example, a constant $\omega$ as used by Brans-Dicke \cite{brans:1961fk} does not satisfy this condition. Such a condition is respected for coupling functions of the form $2\omega+3 \propto (\Phi^n-\Phi_\s^n)^{-1}$ or $2\omega+3 \propto (\Phi-\Phi_\s)^{-n}$, for $n>1$ but is far from being generically satisfied. Another example satisfying (\ref{eq:condDEV}) is given by a coupling function of the form $2\omega+3=-k\left(\ln \Phi/\Phi_s\right)^2$ which is important since the conformal scale factor to transform the action (\ref{eq:actiondila}) into the Einstein frame is then given by $B(\varphi)=e^{k\phi^2/2}$ (where $\varphi$ is a rescaled scalar field) which is widely considered in the literature (see also the discussion in Sec.~\ref{sec:nonperturb}). 

Therefore, one should note that a wide range of theories usually considered in the literature satisfy such a condition \cite{garciabellidoPLB90,sernaPRD96,barrowPRD97,sernaCQG02}. Theories such that equation (\ref{eq:condDEV}) is not respected cannot be treated with the current perturbative approach. Although it restricts the approach originally developed in \cite{jarvPRD12}, the outcome of such a study is nevertheless very informative on the kind of behaviours that scalar-tensor theories can have in the so-called GR limit.

\subsection{Damour and Polyakov's dilaton}
Let us now examine specific cases of the general action (\ref{eq:actiondila}). First of all, let us consider the string dilaton considered in \cite{DamPolyGRG94}. For this class of theory, the parameter $n$ takes a value of unity (see Appendix \ref{app:dilaton}) and the critical value $D=1$ therefore translates into $A_\s \Phi_\s = 3/8$ (while the critical value for standard scalar-tensor theories characterized by $n=0$ is given by $A_\s\Phi_\s=-3/8$). Therefore the dilaton considered in \cite{DamPolyGRG94} can have both convergent and divergent behaviours depending on the function $\omega(\Phi)$. If the function $\omega$ is such that $A_\s\Phi_\s<1$, the dilaton will converge towards GR while it will (locally) diverge in the other cases (see also Sec. \ref{sec:nonperturb}).

\subsection{Pressuron without potential}
\label{sec:pressuron}

In \cite{nousPRD13}, we recenlty showed that the massless pressuron is not constrained by current solar system observations because of the intrinsic decoupling occurring when $n=1/2$ in pressure-less regimes. In particular, the coupling function $\omega(\Phi)$ is weakly constrained in the case of a pressuron ($\omega \sim 1$ is still allowed by solar system observations while one needs $\omega > 10^4$ for usual Brans-Dicke theories \cite{Brans:2014}). Therefore, the GR limit assumptions are not all justified  for such a class of theory. Nevertheless, we shall investigate the pressuron's behaviour in the GR limit in order to compare it with the dilaton and usual generalized Brans-Dicke cases \footnote{\emph{generalized} here means $\omega \rightarrow \omega(\Phi)$.}. After that, we shall relax assumption (a) and study the case where $\omega(\Phi)$ is finite. However, constraints on the apparent non-variation of the fundamental constants seem to indicate that the scalar-field, if it exists, must be close to a constant during the visible epoch. Therefore, assumption (b) is still required in order to explain the observed constancy of the fundamental constants.

\subsubsection{GR limit}
The pressuron case is singular in the sense that $D=1$ in any case, while it satisfies the necessary conditions imposed by the GR limit $A_\s\neq 0$ and $\Phi_\s \neq 0$ \cite{jarvPRD12}. The solutions for $D=1$ can be written from (\ref{eq:solDpos})
\begin{subequations}\label{eq:freezpressuronGR}
\bea
\pm x(t)= \left(M_1- \frac{M_2}{t} \right)^2,\\
\pm  h(t)= \frac{2}{3t^2}\left(M_3+ \frac{M_1M_2}{ \Phi_\s} \ln t+ \frac{2M_2^2b}{t}\right) , 
\eea
\end{subequations}
where $M_i$  are integration constants, while $b$ is a constant characterizing the underlying theory given by (\ref{eq:abc}). Therefore, the pressuron theory converges towards GR for any function $\omega(\Phi)$. This convergence is not surprising since the pressuron decouples to matter in pressure-less regimes \cite{nousPRD13}. But the relation (\ref{eq:freezpressuronGR}) shows how the pressuron freezes after entering in the dust regime.


\subsubsection{Relaxing assumption (a)}\label{sec:pressuronrelax}

While solar system constraints impose usual scalar-tensor theories as well as dilaton-like theories to satisfy assumption (a), the pressuron is not subject to this constraint thanks to the post-Newtonian decoupling studied in \cite{nousPRD13}. Therefore, one can relax this assumption and study a more general scenario\footnote{Note that if we relax this assumption, we can also relax the assumptions (c) and (d) that are no longer needed.}. Nevertheless, the constancy of the fundamental constants of Nature seems to indicate that the derivative of the scalar field has to be very small. Therefore, let us develop our system of equations around any given constant scalar field value $\Phi(t)=\Phi_\c > 0$ (which is solution of the Klein-Gordon equation (\ref{eq:cosmoDecP})), such that $\omega(\Phi_\c)$ is non-singular. In other words, we are still considering the assumption (b): $\dot \Phi \sim 0$. We can develop our field equations around the solution where the scalar field is constant
\be \label{eq:expCarre}
\Phi(t)=\Phi_\c+x(t), ~~~~H(t)=H_\c(t)+h(t),
\ee
where $\Phi_\c$ is constant and $H_\c(t)$ is the solution of Eq. (\ref{eq:cosmoDecH}) with a constant scalar field. The equation at zero order for $H_\c(t)$ is given by
\begin{equation}
	\p{H}_\c(t)=-\frac{3}{2} H_\c(t),
\end{equation}
which is the same equation as in GR. The solution of this equation is given by $H_\c(t)=H_{\rm GR}(t)=H_\s(t)$ where $H_\s(t)$ is given by (\ref{eq:HS}). At first order of perturbations, the Eqs. (\ref{eq:cosmo_decoup}) become
\begin{subequations}\label{eq:pressuronPert}
\bea
\pp{x}(t)&=&-3 \p{x}(t) H_\c(t), \\
\p{h}(t)&=&-3H_\c(t) h(t) + \frac{1}{2} H_\c(t) \frac{\p{x}(t)}{\Phi_\c}.
\eea
\end{subequations}
The solutions therefore write:
\begin{subequations}
\be
x(t)=C_1+\frac{C_2}{t},
\ee
and
\be
h(t)=\frac{C_3}{t^2}- \frac{C_2}{3 \Phi_\c} \frac{\ln t}{t^2},
\ee
\end{subequations}
where the $C_i$ are integration constants. This means that the pressuron freezes in the dust regime independently of the initial conditions. This convergence is totally independent of the function $\omega(\Phi)$. The decoupling mechanism of the pressuron is therefore very powerful since the scalar field naturally converges towards a constant scalar field and thus naturally satisfies solar system tests of gravity for any non-singular function $\omega(\Phi)$.

\subsection{General case with a potential in the GR limit}
The analysis done so far does not consider any self-interacting potential for the scalar field.  This potential is phenomenologically motivated if one wants to explain the acceleration of the cosmic expansion (as we will show in Sec. \ref{sec:obs}). Therefore, we will extend the results presented previously by including a potential.

In this section, we will consider the so-called GR limit using the assumptions (a)-(d) presented in Sec.~\ref{sec:pertGR}. Once again, we expand the field equations (\ref{eq:cosmo_decoup}) around the GR limit using (\ref{eq:expansion}). The zeroth order equation then writes
\be \label{eq:zeroP}
\p{H}_\s(t)=- \frac{3}{2} H^2_\s(t)+\frac{V_\s}{4\Phi_\s},
\ee
with $V_\s=V(\Phi_\s)$ and where this is the standard GR equation with $V_\s/2\Phi_\s$ identified with the cosmological constant. 

The solution is
\be
H_\s(t)= \sqrt{\frac{V_\s}{6\Phi_\s}} \tanh \left(\sqrt{\frac{3 V_\s}{8\Phi_\s}} t + K \right), \label{eq:Hbignonassympt}
\ee
 where $K$ is a constant of integration. This is the standard GR solution for a universe with matter density and a cosmological constant which tends towards a de-Sitter space-time characterized by a constant Hubble rate
\begin{equation}\label{eq:Hassympt}
	H_\s(t\rightarrow \infty)=H_{\s\infty}=\sqrt{\frac{V_\s}{6\Phi_\s}}.
\end{equation}

The first perturbative order equations write:
\begin{subequations}\label{eq:limitGRPot}
\bea
\pp{x}(t)&=&-3 H_\s(t)\p{x}(t)+\frac{\p{x}^2(t)}{2x(t)}+3(1-2n)A_\s\Phi_\s H^2_\s(t) x(t) \nonumber \\
&+& \left(W_\s -\frac{1-2n}{2}V_\s\right) A_\s x(t),  \label{eq:xwP}
\eea
 and
\bea
\p{h}(t)&+&3H_\s(t)h(t)= -\frac{1}{4 \Phi_\s} \left(1+\frac{1}{2A_\s \Phi_\s} \right) \frac{\p{x}^2(t)}{x(t)} \nonumber\\
&+& \frac{1}{2\Phi_\s} H_\s(t) \p{x}(t) -\frac{3}{2}(1-2n)A_\s H_\s^2(t)x(t) \nonumber\\
&+&\left(\tilde W_\s+\frac{1-2n}{4}\frac{ V_\s}{\Phi_\s}\right)  A_\s x(t),
\eea
\end{subequations} 
with  $V_\s'=dV/d\Phi(\Phi_\s)$ and
\begin{subequations}
	\begin{eqnarray}
		W_\s&=&2V_\s-\Phi_\s V_\s'  \label{eq:W}\\
		\tilde W_\s&=&\frac{\Phi_\s V'_\s-V_\s}{4A_\s \Phi^2_\s} -\frac{ W_\s}{2}  \\
		&=& \frac{V_\s'}{4A_\s \Phi_\s}(1+2A_\s\Phi_\s^2)-\frac{V_\s}{4A_\s \Phi_\s^2}(1+4A_\s\Phi_\s^2)\, .\nonumber
	\end{eqnarray}
\end{subequations}
In general, it is not possible to find an analytical solution of the Eq. (\ref{eq:limitGRPot}). It is nevertheless still possible to study the asymptotic behaviour of these equations. The asymptotic behaviours of Eq. (\ref{eq:limitGRPot}) is obtained by replacing $H_\s(t)$ by its asymptotic expression given by (\ref{eq:Hassympt}):
\begin{subequations}\label{eq:asymptotic}
	\begin{eqnarray}
		\pp{x}(t)&\infeq &-3 H_{\s\infty} \p{x}(t)+\frac{\p{x}^2(t)}{2x(t)}+A_\s W_\s x(t) \\
\p{h}(t)&+&3H_{\s\infty}h(t)\infeq	-\frac{1}{4 \Phi_\s} \left(1+\frac{1}{2A_\s \Phi_\s} \right) \frac{\p{x}^2(t)}{x(t)} \nonumber\\
	&+& \frac{1}{2\Phi_\s} H_{\s\infty} \p{x}(t) +\tilde W_\s  A_\s x(t).
	\end{eqnarray}
\end{subequations}
It is remarkable that, due to an exact cancellation of several terms, these asymptotic equations are now completely independent of $n$ --- which means that the form of the coupling function $f(\Phi)$ does not have any influence on the asymptotical cosmological evolution. This is explained by the fact that asymptotically, the influence of the potential will always be stronger than the influence of the matter density $\rho$. Therefore, the system will asymptotically behave as if there is no matter ($\rho=0$). Hence, for $V \neq 0$, it is quite logical that the scalar-to-matter coupling has no influence asymptotically. 

As a consequence, when considering a potential in the action (\ref{eq:actiondila}), the asymptotical solutions are the same as in standard generalized Brans-Dicke theory characterized by $n=0$ (or $f(\Phi)=1$). This case has been studied in \cite{jarvPRD10,jarv:2010fk}. In particular, the solutions depend on a new critical parameter\footnote{This parameter is the same as in \cite{jarv:2010fk}. Notice that the potential $V_{\textrm {JKS}}$ used in \cite{jarv:2010fk} is related to the one used in this paper by a multiplicative constant $2\kappa V_{\textrm {JKS}}=V$.}
\begin{eqnarray}\label{eq:C}
	C&=&	2A_\s W_\s +\frac{3V_\s}{2\Phi_\s}=2A_\s W_\s+C_1\\
	&=&\frac{V_\s}{2\Phi_\s}(3+8A_\s\Phi_\s)-2A_\s V_\s'\Phi_\s \, , \nonumber
\end{eqnarray}
with
\begin{equation}\label{eq:C1}
	C_1=\frac{3V_\s}{2\Phi_\s}\, .
\end{equation}

As mentioned the solutions are the same as in \cite{jarv:2010fk}. Nevertheless, for the sake of completeness, they are recalled in Appendix~\ref{app:generalPot}. Three behaviours can be exhibited (see \cite{jarv:2010fk} for a detailed studies):
\begin{itemize}
	\item $C > 0$ : The solutions are exponential (see \ref{eq:genPotPos}). They are  exponentially  converging towards GR if $A_\s W_\s <0$.
	\item $C = 0$ : The solutions are linear exponential (see \ref{eq:genPot0}).  They converge towards GR.
	\item $C<0$ : The solutions are damped oscillations (see \ref{eq:genPotNeg}). They converge towards GR.
\end{itemize}
These asymptotic solutions applied for the dilaton as well as for the pressuron.

\subsection{Pressuron with a potential}\label{sec:pressuronPot}
In this section, we will study more carefully the case of the pressuron. First, the solutions of the full GR limit equations (\ref{eq:limitGRPot}) take an analytical form given in Appendix \ref{ap:pressuron}. They give a more detailed evolution than the asymptotic behaviour computed from Eq. (\ref{eq:asymptotic}). Nevertheless, the behaviour is similar to what is described in the previous section.

As mentioned in Sec. \ref{sec:pressuronrelax}, the pressuron does not need to satisfy the GR limit because this theory automatically satisfies solar-system tests independently of the function $\omega(\Phi)$ \cite{nousPRD13}. Therefore, we can relax the assumptions (a), (c) and (d) similarly to what was done in Sec. \ref{sec:pressuronrelax} and develop the equations around any given constant scalar field value $\Phi(t)=\Phi_\c > 0$ as it is done in Eq (\ref{eq:expCarre}).
 In order to use such an expansion,  $\Phi(t)=\Phi_\c$ has to be a solution of the zeroth order perturbation of Eq. (\ref{eq:cosmoDecP}). This is only the case for particular potentials satisfying the condition
\be
W_\c=2V_\c-\Phi_\c V'_\c=0 \, , \label{eq:condition0}
\ee
with $V_\c=V(\Phi_\c)$ and $V_\c'=dV/d\Phi (\Phi_\c)$.

Using the perturbation scheme (\ref{eq:expCarre}), Eqs.~(\ref{eq:cosmo_decoup}) at first order can be written
\begin{subequations}\label{eq:pressuronPotPert}
	\begin{eqnarray}
		\ddot x(t)&=&-3H_\c(t)\dot x(t)+\frac{W_\c'}{2\omega_\c+3} x(t)\label{eq:pertXPressPot}\\
		\dot h(t)&+&3H_\c(t)h(t)=\frac{H_\c(t)\dot x(t)}{2\Phi_\c}-\frac{1}{2}\frac{W_\c'}{2\omega_\c+3}x(t)\nonumber\\
		&&\qquad +\frac{V_\c-W_\c}{4\Phi_\c^2}x(t),
	\end{eqnarray}
\end{subequations}
with  $\omega_\c=\omega(\Phi_\c )$ and
\begin{equation}
	W'_\c=V_\c'-\Phi_\c V_\c'' \, .
\end{equation}
Finally, $H_\c(t)$ is the zeroth order solution of Eq. (\ref{eq:cosmoDecP}) which becomes identical to Eq. (\ref{eq:zeroP}). The solution of this equation is given by Eq. (\ref{eq:Hbignonassympt}) if $V_\c\neq 0$ and  by (\ref{eq:HS}) when $V_\c=0$.

The condition  (\ref{eq:condition0}) can be satisfied in different manners:
\begin{enumerate}[(i)]
	\item $2V_\c=\Phi_\c V'_\c \neq 0$: this is the case for $V(\Phi)=a\Phi^2$ for any $\Phi_\c\neq 0$ which is a little bit particular since $W_\c'=0$ too. Note that for every function $g(\Phi)$, the potential defined by $V(\Phi)=g(\Phi)-g(\Phi_\c)+\frac{\Phi_\c}{2}g'(\Phi_\c)$ will satisfy this condition.
	
	In this case, the solution of the zeroth order equation gives (\ref{eq:Hbignonassympt})

		\begin{equation}
			H_\c(t)=\sqrt{\frac{V_\c}{6 \Phi_\c}}\tanh \left(\sqrt{\frac{3V_\c}{8\Phi_\c}t}\right)\infeq\sqrt{\frac{V_\c}{6 \Phi_\c}},
		\end{equation}
		where the integration constant $K$ has been set up to 0 (which corresponds to a redefinition of the time origin). The solution of Eqs. (\ref{eq:pressuronPotPert}) depends on a new critical parameter 
		\begin{equation}\label{eq:B}
			B=4\frac{W'_\c}{2\omega_\c+3}+\frac{3V_\c}{2\Phi_\c}=4\frac{W'_\c}{2\omega_\c+3}+C_1 \, ,
		\end{equation}
		with  $C_1$ given by (\ref{eq:C1}).
		
		For the sake of conciseness, the solutions are developed in details in Appendix~\ref{app:pressuronPot}.
		
		\begin{itemize}
			\item $B > 0$ : The solutions are exponential (see Eqs.~(\ref{eq:pressuronPotPos})).	In particular, the solution will converge towards a constant scalar field only if $\frac{W'_\c}{2\omega_\c+3} \leq 0$ (which is the case of a quadratic potential). In the other cases, the solution will diverge. 
	
	\item $B=0$ or $\frac{W'_\c}{2\omega_\c+3}=-\frac{3V_\c}{8\Phi_\c}$ : The solutions are linear exponential (see Eqs.~(\ref{eq:pressuronPot0})). They always converge towards a constant scalar field.
	
	\item $B<0$ : The solutions are damped oscillating (see Eqs.~(\ref{eq:pressuronPotNeg})).
They always converge towards a constant scalar field.
		
\end{itemize}
In conclusion for this first case, the solutions will diverge exponentially if $\frac{W_\c'}{2\omega_\c'+3}>0$ and will converge in all the other cases: exponentially if $0\geq \frac{W_\c'}{2\omega_\c'+3} \geq -\frac{2V_\c}{3\Phi_\c}$ and following damped oscillations if $-\frac{2V_\c}{3\Phi_\c}>\frac{W_\c'}{2\omega_\c'+3}$.

	\item $V_\c=V_\c'=0$ and $W_\c'=-\Phi_\c V_\c''\neq 0$: this is the case of quadratic  and quartic potential which can be interesting in the context of Higgs field \cite{fuzfa:2013vn}. In fact any potential of the form $V(\Phi)=\sum_{i>1}\alpha_i(\Phi-\Phi_\c)^i$ would satisfy the mentioned conditions.
	
	In this case, the solution of the zeroth order equation gives (\ref{eq:HS})
	\begin{equation}
		H_\c(t)=\frac{2}{3t}
	\end{equation}
	where the integration constant has been set up to 0 for convenience. The solution of Eqs. (\ref{eq:pressuronPotPert}) depend on a critical parameter that is simply $\frac{W_\c'}{2\omega_\c+3}=-\frac{\Phi_\c V_\c''}{2\omega_\c+3}$ 
	\begin{itemize}
		\item $\frac{W'_\c}{2\omega_c+3}>0$ or $\frac{V_\c''}{2\omega_c+3}<0$ : The solutions are exponentially divergent  (see Eq.~(\ref{eq:pressuronPot2Pos})).
			
		\item $\frac{W'_\c}{2\omega_c+3}< 0 $	or $\frac{V_\c''}{2\omega_c+3}>0$ : The solutions are  damped oscillations  (see Eq.~(\ref{eq:pressuronPot2Neg})). They always converge towards a constant scalar field.

	\end{itemize}
In conclusion, if $\frac{V_\c''}{2\omega_c+3}>0$, the solution will always converge towards a constant scalar field.
	
	\item $V_\c=V_\c'=V_\c''=W'_\c=0$: in this case, the perturbed equations (\ref{eq:pressuronPotPert}) are similar to the ones without potential (\ref{eq:pressuronPert}). This means the potential is too smooth to have any influence at first order. The perturbative approach used here is not informative in this case and the cosmological behaviour is similar to the one developed in Sec. \ref{sec:pressuronrelax}.
\end{enumerate} 

Finally, let us stress that the perturbative approach is limited as one can see with equation (\ref{eq:condition0}). Thus, one should not be too conclusive with respect to the massive pressuron stability/convergence at this stage. However, the present result seems to indicate that the cosmological evolution of the pressuron seems to be very stable since it converges towards a constant value for a large class of cases. Further non-perturbative approach should be considered in order to figure this out.

\subsection{Non-perturbative result}
\label{sec:nonperturb}

The results presented up to now rely on a perturbation scheme. In particular, we always study the case where the GR limit is valid (or in the case of the pressuron where the slow variation of the scalar field is valid). In this section, we present an analytical way to treat the problem of the convergence of the scalar field  in a non-perturbative way in the case where no potential is present. The procedure is inspired from \cite{sernaCQG02}. It corresponds to study the evolution of the scalar field in the so-called Einstein frame where the gravitational part of the action take the same form as in GR. This frame, obtained by a conformal transformation can be useful in particular to study the convergence of the scalar field. Nevertheless, the GR limit is not defined in this frame. The full conformal transformation is explicitly detailed in \cite{moiPRD13_2}.

We define the parameter $p=\ln \left(\sqrt{\Phi} a \right)$ as in \cite{sernaCQG02}. Then, from Eqs.~(\ref{eq:field1}), one can derive a decoupled scalar-field equation that writes (the detailed calculations are similar to the ones in \cite{sernaCQG02})
\be
\frac{2 \left(W^{1/2} \psi' \right)'}{1-W \psi'^2}+3(1-w) \psi' W^{1/2} = \frac{1-3w-2n}{W^{1/2}}, \label{eq:decoupledJF}
\ee
where 
\begin{subequations}
	\begin{eqnarray}
		\psi&=&\frac{1}{2}\ln\Phi, \\
		W(\psi)&=&\frac{3+2\omega(\Phi)}{3}=\frac{3+2\omega(e^{2\psi})}{3},
	\end{eqnarray}
\end{subequations}
$w=P/\epsilon$ and $X' \equiv d X / dp$. One can see that the source term vanishes for $n=1/2$ (pressuron) and $w = 0$ (dust matter). 

A conformal transformation ${g}_{\mu \nu}=\Phi^{-1} g^*_{\mu \nu}=B^2(\varphi)g^*_{\mu\nu}$ where $B$ is a conformal factor depending from a rescaled scalar field $\varphi$, allows to put the last equation in a more convenient form by working in the Einstein representation\footnote{The stars indicate quantities expressed in the Einstein frame.}. The new rescaled scalar field $\varphi$ is defined from the differential relation
\begin{equation}\label{eq:alpha}
\alpha(\varphi)=\frac{\partial \ln B(\varphi)}{\partial \varphi}=-\frac{1}{2}\frac{\partial \ln \Phi}{\partial \varphi}=-\frac{\partial \psi}{\partial \varphi}=\frac{1}{\sqrt{3+2\omega}}	.
\end{equation}	
The insertion of the last expression into (\ref{eq:decoupledJF}) leads to the following equation
\be\label{eq:varphi}
\frac{2 \varphi''}{3- \varphi'^2} +\left(1-w \right)  {\varphi}'= - \left(1-2n-3 w \right) \alpha(\varphi).
\ee
This equation is a generalization of the one found in \cite{sernaCQG02} recovered when $n=0$ and is also in agreement with the developments done in \cite{damourPRD93}. This equation is exact and gives the evolution of the rescaled scalar field $\varphi$ as a function of the $p$ variable.

The solution for a pressuron ($n=1/2$) in the matter era ($w= 0$) is given by an exponential damping and writes 
\be 
{\varphi}(p) = \varphi_\infty \pm \frac{2}{\sqrt{3}} \ln \left[K e^{-\frac{3}{2}p}+ \left(1+K^2 e^{-3p} \right)^{1/2} \right],
\ee
where ${\varphi}_\infty$ is the constant value of ${\varphi}$ at $p \rightarrow \infty$ and $K$ is a constant of integration depending on the  initial conditions
\begin{equation}
	K=\frac{\varphi_0'}{\sqrt{3-\varphi_0'^2}} \, .
\end{equation}

This non-perturbative result confirms that the pressuron converges and tends to a constant in any case in the matter era, independently of the initial conditions or of the function $\omega(\Phi)$. 

The dilaton characterized by $n=1$ will have a convergence mechanism depending on the coupling function \cite{minazzoli:2013ve}. An example often used consists to consider $B(\varphi)=e^{k\varphi^2/2}$ which leads to $\alpha=k\varphi$ and corresponds to $3+2\omega(\Phi)=-\left(k\ln \frac{\Phi}{\Phi_0}\right)^{-1}$. The integration of (\ref{eq:varphi}) in the non-relativistic limit (ie. $3- \varphi'^2 \rightarrow 3$ \cite{damourPRL93,*damourPRD93}) shows that the behaviour of the scalar field depends on a critical parameter $D=1+8k/3$ which corresponds to  the critical parameter (\ref{eq:D}). We have three different regimes that exactly corresponds to the solutions presented in Sec.~\ref{sec:pertGR}. In particular, the convergence towards GR appears when $D<1$ or when $k<0$. A positive value of $k$ leads to divergent scenarios only.


\section{Observables}\label{sec:obs}
As mentioned in the introduction, since the coupling between the scalar field and matter is non-minimal, the observables are not necessary derived in the same way as in GR. In this section, we derived two types observables related to the late time cosmological evolution of the theory: the temporal variation of the fundamental constants and the distance-luminosity versus redshif relation. We derive these observables from first principles. Afterwards, we use the evolutions derived in Sec.~\ref{sec:evol} to compare quantitatively the predictions with the observations.

\subsection{Time variation of the fundamental coupling constants}

\subsubsection{The fine structure constant}

The fine-structure constant is the one for which its time variation is the best constrained \cite{fujiLNP04, *petrovPRC06,*guena:2012hl}. According to the general action (\ref{eq:actiondila}), the fine structure constant $\alpha = e^2/ \hbar c$ is proportional to $f^{-1}$ \cite{DamPolyGRG94,damourPRD02} such that
\be
\frac{\dot{\alpha}}{\alpha}= - \left.\frac{\dot{f}}{f}\right|_0= -n \left.\frac{\dot{\Phi}}{\Phi}\right|_0= -n \frac{\dot{x_0}}{\Phi_\star},
\ee
 where the subscript $0$ indicates that we deal with values at present epoch and $\Phi_\s$ is the asymptotic value of the scalar field\footnote{Note that in the case of the pressuron, we have denoted $\Phi_\s$ by $\Phi_\c$.} and $x(t)$ is the first order solution which have been developed in details in the previous section.
The constraints on the temporal variation of the fine structure constant \cite{damourNPB96, fujiLNP04, *petrovPRC06,*guena:2012hl} therefore gives a constraint which can be written
\begin{equation}\label{eq:alphaDot}
\left|\frac{\dot{x}_0}{\Phi_\star}\right| \lesssim 10^{-16} ~yr^{-1}.	
\end{equation}
This impressive constraint seems to favor the behaviours converging towards a constant scalar field in the zoo of all the solutions developed in the previous section.

\subsubsection{The gravitational constant}
\label{sec:effect_grav_const}

Scalar-tensor theories generically predict a variation of the effective constant of gravitation. Such an effective constant appears in the Poisson equation at the zeroth order in the post-Newtonian perturbative development of the theory. It depends on the cosmological background value of the scalar-field. The effective gravitational constant for general universal multiplicative coupling is given by \cite{moiPRD13_2}
\be
G_{\textrm {eff}} =\frac{c^4\kappa}{8 \pi} \left(1+\frac{1-2n}{2 \omega_0+3} \right)  \frac{f(\Phi_0)}{\Phi_0}.
\ee
Therefore, the time derivative of this expression leads to
\begin{eqnarray}
	\frac{\dot G_{\textrm {eff}}}{G_{\textrm {eff}}}&=&(n-1)\frac{\dot \Phi_0}{\Phi_0}-\frac{1-2n}{(\omega_0+2-n)(2\omega_0+3)}\omega_0'\dot \Phi_0\nonumber\\
	&=&(n-1)\frac{\dot \Phi_0}{\Phi_0}+\frac{2\omega_0+3}{2\omega_0+4-2n}(1-2n)A_0\dot \Phi_0,
\end{eqnarray}
where $A(\Phi)$ is defined by (\ref{eq:defA}).

In particular, the specific case of the pressuron ($n=1/2$) gives
\be
G_{\textrm{eff}} =\frac{c^4\kappa}{8 \pi} \frac{1}{\sqrt{\Phi_0}}.
\ee
In this last case, the variation of the gravitational constant is given by
\be
\frac{\dot{G}_{\textrm{eff}}}{G_{\textrm{eff}}} = - \frac{1}{2} \frac{\dot{\Phi}_0}{\Phi_0}= - \frac{1}{2}\frac{\dot{x}_0}{\Phi_\star}.
\ee
Hence, lunar laser ranging constraint on the variation of the gravitational constant  \cite{williams:2009ys} gives the following constraint on the pressuron cosmological perturbation at present epoch
\be\label{eq:GdotPressuron}
\left| \frac{\dot{x}_0}{\Phi_\star} \right| = (8 \pm 18) ~10^{-13}\,  yr^{-1}.
\ee
On the other hand, for Damour \& Polyakov's dilaton \cite{DamPolyGRG94} ($n=1$), one has
\be
G_{\textrm{eff}}  =\frac{\kappa c^4}{8 \pi} \left(\frac{2 \omega_0+2}{2 \omega_0 +3} \right) 
\ee
and
\be
\frac{\dot{G}_{\textrm{eff}}}{G_{\textrm{eff}}}  =   \frac{2 \omega_0+3}{2 \omega_0+2} ~A_0 ~ \dot{\Phi}_0.
\ee
Given the fact that equivalence principle violation constraints (from composition-dependent effects) are pretty strong on a massless dilaton ($\omega_0 > 10^{10}$) \cite{damourPRD10},  lunar laser ranging constraint on the gravitational constant variability gives:
\be\label{eq:GdotDilaton}
\frac{\dot{G}_{\textrm{eff}}}{G_{\textrm{eff}}}  \sim  A_0 ~\dot{x}_0 = (4 \pm 9)\times\, 10^{-13} yr^{-1}.
\ee

From the Eqs. (\ref{eq:alphaDot}), (\ref{eq:GdotPressuron}) and (\ref{eq:GdotDilaton}), we can see that the constraint on the fine structure constant is the one who gives the stringent constraint on the present value of the derivative of the scalar field\footnote{Note that in the case of a standard generalized Brans-Dicke theory ($n=0$), there is no variation of the fine structure constant. Therefore, the constraint on the variation of the gravitational constant is important (see \cite{damour:1990fk})}. While the constraint (\ref{eq:alphaDot}) does not strictly exclude solutions where the scalar field does not converge towards a constant (one can imagine an unlikely scenario where the divergence is very slow), this is a strong indication that the scalar field needs to converge in the late-time cosmological evolution. 

Let us remind that the pressuron ($n=1/2$) converges towards a constant in the matter era independently of the function $\omega(\Phi)$. The theories characterized by other coupling functions converge if $D<1$ which is equivalent to 
\begin{equation}
	(1-2n)A_\s\Phi_\s < 0.
\end{equation}


\subsection{Distance-luminosity and Supernovae Ia data}
In this section, we will show how to compute the distance-luminosity from the action (\ref{eq:actiondila}). The procedure is similar to what is done in GR in \cite{ellis:2012}. In order to derive the distance-luminosity relation, we have to determine how light propagates in the theory parametrized by the action (\ref{eq:actiondila}). Introducing the electromagnetic Lagrangian in the action (\ref{eq:actiondila}) and varying this action with respect to the 4-potential $A^\mu$ leads to modified Maxwell equations. In a vacuum, these equations reduce to
\begin{equation}\label{eq:maxwell}
	\nabla_\nu \left(f(\Phi)F^{\mu\nu}\right)=0
\end{equation}
where $F^{\mu\nu}=A^{\nu,\mu}-A^{\mu,\nu}$ is the standard Faraday tensor. Following the analysis made in \cite{MTW}, we expand the four-vector potential as 
\begin{equation}\label{eq:expansionEM}
A^\mu = \Re \left\{ \left(b^\mu + \epsilon c^\mu + O(\epsilon^2) \right) \exp^{i \theta / \epsilon} \right\} \ .	
\end{equation}

The introduction of this expansion in (\ref{eq:maxwell}) and the use of the Lorenz gauge lead to the usual null-geodesic equation at the geometric optic limit (see \cite{nousPRD13,moiPRD13}). The next-to-leading order of the modified Maxwell equations (see the procedure used in \cite{MTW,moiPRD13}) is given by
\begin{subequations}
	\begin{eqnarray}
		k^\nu \nabla_\nu b &=&-\frac{1}{2}b\nabla_\nu k^\nu -\frac{1}{2}bk^\nu \partial_\nu \ln f(\Phi) \label{eq:amplitude}\\
		k^\nu \nabla_\nu h^\mu &=&\frac{1}{2}k^\mu h^\nu\partial_\nu \ln f(\Phi)
	\end{eqnarray}	
\end{subequations}
 \footnote{Note that there is a typo in (30) in \cite{moiPRD13}.} where $b$ is the amplitude of $b^\mu$, $h^\mu$ is the polarisation vector given by $b^\mu=b h^\mu$ and $k_\mu \equiv \partial_\mu \theta$. From there, it follows that the conservation law of the number of photons (or intensity) is modified:
\begin{equation}
	\nabla_\nu \left(b^2k^\nu\right)=-b^2 k^\nu\partial_\nu \ln f(\Phi).
\end{equation}
Now, let's take a radial light ray emitted in coordinates $(ct_{e},r_e=0,0,0)$ (in spherical coordinates) and observed in coordinates $(ct_0,r_0,0,0)$. The coordinates of the wave vector are given by 
$$k^\mu=dx^\mu/d\lambda=(k^0,k^r,0,0)=(-k_0,k_r/a^2(t),0,0)$$
 with $\lambda$ an affine parameter on the null geodesic. The fact that the wave vector is a null vector implies that $k_r=a(t)k_0$. Since the metric is independent of the radial coordinate, the geodesic equation tells us that $k_r$ is conserved. Finally, let us notice that
\begin{equation}\label{eq:dtdl}
	\frac{dt}{d\lambda}=k^0=-k_0=-\frac{k_r}{a(t)}.
\end{equation}
The equation of the amplitude of the electromagnetic signal (\ref{eq:amplitude}) in flat FLRW geometry for a radial light ray can be written as
\begin{equation}
	\frac{db}{d\lambda}+\frac{b}{2}\frac{1}{r^2a^3(t)}\frac{d\left(r^2a^3(t)k^0\right)}{dt}+\frac{b}{2}\frac{d \ln f(\Phi)}{d\lambda}=0
\end{equation}
Using the fact that $k^0=-k_r/a(t)$, the fact that $k_r$ is constant on the light ray trajectory and (\ref{eq:dtdl}), the last equation becomes
\begin{equation}
	\frac{d\ln b}{d\lambda}+\frac{1}{2}\frac{d \ln r^2 a^2(t)}{d\lambda}+\frac{1}{2}\frac{d \ln f(\Phi)}{d\lambda}=0,
\end{equation}
which means the quantity $K=b(t,r)ra(t)\sqrt{f(\Phi(t))}$ is constant along the light ray.

The flux of energy measured by an observer is given by \cite{MTW,ellis:2012}
\begin{equation}
	F_0=\left | u_\mu n_\nu T^{\mu\nu}\right|
\end{equation}
where $u_\mu$ is the 4-velocity of the observer ($u_\mu=(-1,0,0,0)$ for a static observer), $n_\nu$ is a unit vector pointing in the direction of the light ray ($n_\nu=(0,a(t),0,0)$ in case of a radial light propagation) and $T^{\mu\nu}$ is the standard stress-energy tensor for electromagnetism:
$$
T^{\mu\nu}=F^{\mu\alpha}F^{\nu}_{\ \alpha}-\frac{1}{4}g^{\mu\nu}F^{\alpha\beta}F_{\alpha\beta},
$$
which gives at leading order:
\be
T^{\mu\nu}=\Re \left\{ i  e^{i\theta/\epsilon}~b^2k^\mu k^\nu\right\},
\ee
using the expansion (\ref{eq:expansionEM}).
The flux of energy measured is thus given by
\begin{equation}\label{eq:flux_tmp}
 F_0 =\left|a_0 b^2 k^0 k^r\right| =\frac{k_r^2b^2 }{a^2_0}=\frac{k_r^2 K^2}{r_0^2 a^4_0f(\Phi_0)}=\frac{C}{r_0^2a_0^4f(\Phi_0)}
\end{equation}
where $C$ is a constant over the null geodesic and indices $0$ refer to the measurement (made at present epoch). A similar expression can be computed for the emitted flux
$$
F_e=\frac{C}{r_e^2a_e^4f(\Phi_e)}
$$
where indices $e$ refer to the emission of the signal. The angular integral of this emitted flux gives the emitted luminosity $L_e$
$$
L_e=\frac{4\pi C}{a_e^2f(\Phi_e)}.
$$
Also, the expression of the distance-luminosity is defined by 
$$
	d_L=\left(\frac{L_e}{4\pi F_0}\right)^{1/2}=\frac{a_0}{a_e}a_0 r_0 \sqrt{\frac{f(\Phi_0)}{f(\Phi_e)}}.
$$
Finally, using $ds^2=0$ and integrating over the null-geodesic, we get $r_0=c\int_{t_e}^{t_0}\frac{dt}{a}=\frac{c}{a_0}\int_0^z\frac{dz}{H(z)}$ where $z$ is the redshift defined as $1+z=\frac{\nu_e}{\nu_0}=\frac{a_0}{a_e}$. The last equation can then be written as\footnote{Note that a similar expression can be derived in curved FLRW space-time following the same reasoning.}

\begin{equation}
	d_L=c(1+z)\sqrt{\frac{f(\Phi(z=0))}{f(\Phi(z))}}\int_0^z\frac{dz}{H(z)}.
\end{equation}
If we introduce the conformal time $\eta$ define by $dt=a/a_0 d\eta$, we get
\begin{eqnarray}
	d_L&=&c(1+z)(\eta_0-\eta)\sqrt{\frac{f(\Phi(\eta_0))}{f(\Phi_(\eta))}}=c \frac{a_0}{a}(\eta_0-\eta)\sqrt{\frac{f(\Phi(\eta_0))}{f(\Phi_(\eta))}}\nonumber\\
	&=&c\frac{\tilde a_0}{\tilde a}(\eta_0-\eta)\label{eq:distLum}
\end{eqnarray}
where in the last expression, we introduced an effective cosmic scale factor $\tilde a=a \sqrt{f(\Phi)}$. First of all, this expression reduces to the standard GR expression when $f(\Phi)=1$. As we can see, SNe Ia data's are sensitive to the evolution of $\tilde a$ and not to the evolution of the cosmic scale factor $a$. This means it might be conceivable to have an acceleration of the effective cosmic scale factor $\tilde a$ able to reproduce SNe data, while the cosmic scale factor $a$ is not accelerated. In order to study such a possibility, let us define 
\begin{equation}
	\tilde H=\frac{\dot {\tilde a} }{\tilde a}=H+\frac{1}{2}\frac{f'(\Phi)}{f(\Phi)}\dot \Phi=H+\frac{n}{2}\frac{\dot \Phi}{\Phi} \ ,
\end{equation}
and introduce the perturbative results obtained in section \ref{sec:evol}. By denoting $H_{\rm GR}=H_\s=H_\c$, we have
\begin{equation}\label{eq:Htilde}
	\tilde H=H_{GR}(t)+h(t)+\frac{n}{2}\frac{\dot x(t)}{\Phi_\s}.
\end{equation}
The last term comes from the fact that the coupling between the scalar field and matter is not minimal (in particular, it vanishes for minimal coupling characterized by $n=0$). 

In the same spirit, we can defined the observed acceleration parameter by
\begin{equation}\label{eq:qtilde}
	\tilde q=\frac{\ddot{\tilde a} }{\tilde a \tilde H^2}=1+\frac{\dot {\tilde H}}{\tilde H^2}.
\end{equation}
Inserting (\ref{eq:Htilde}), the acceleration parameter is given at first order by
\begin{equation}
	\tilde q=q_{GR}(t) +\frac{\dot h(t)}{H_{GR}^2(t)}+\frac{n}{2}\frac{\ddot x(t)}{\Phi_\s H^2_{GR}(t)}
\end{equation}
where $q_{GR}=1+\frac{\dot H_{GR}}{H_{GR}^2}$ is the acceleration parameter obtained in GR. The observed acceleration parameter needs to be positive in order to explain the apparent acceleration of the cosmic expansion as observed by SNe Ia data. We can distinguish two cases amongst the solutions studied in the previous section:
\begin{itemize}
	\item if $H_{GR}$ is given by (\ref{eq:HS}). This case appears when no potential is considered in the action (\ref{eq:actiondila}) and in the case of the pressuron ($n=1/2$)  with a potential characterized by $V(\Phi_\c)=V'(\Phi_\c)=0$ and $W_\c\neq 0$ (see the case (ii) from Sec.~\ref{sec:pressuronPot}). In this case, $q_{GR}=-1/2$. The observed acceleration parameter is then given by
	\begin{equation}
		\tilde q=-\frac{1}{2}+\frac{9t^2}{4}\left(\dot h(t)+\frac{n}{2}\frac{\ddot x(t)}{\Phi_\s}\right).
	\end{equation}
In order to have $\tilde q >0$ for large value of $t$, the term into parenthesis needs to be asymptotically larger than $1/t^2$. This means that asymptotically, $x(t)$ needs to be larger than $\ln t$ and $h(t)$ needs to be larger than $1/t$. The solutions developed in Sec.~\ref{sec:evol} show that these conditions are satisfied only for theories where the scalar field does not converge towards a constant. Therefore, the modification of the expression of the distance-luminosity (\ref{eq:distLum}) does not allow one to explain the acceleration of the cosmic expansion for theories where the scalar field converges towards a constant --- which is required from the constraints on the variation of the fundamental coupling constants.

\item if $H_{GR}$ is given by (\ref{eq:Hbignonassympt}). This case appears for theories with potential. In this case, $q_{GR}$ is given by
\begin{equation}
	q_{GR}(t)=1+\frac{2}{3\sinh^2 \left(\sqrt{\frac{3V_\s}{8\Phi\s}}t+K\right)}
\end{equation}
which asymptotically tends towards 1. In this case, the behaviour of the perturbation in (\ref{eq:qtilde}) is not important since the acceleration of the cosmic expansion is produced by the potential that plays the role of a cosmological constant. 
\end{itemize}
Therefore, in the class of theories presented in this paper, the late time cosmic expansion can only be produced by a potential as soon as one wants to also satisfy the constraints on the temporal variations of the constants of Nature.

\section{Conclusion}
\label{sec:concl} 
In this paper, we have studied the late-time cosmological evolution (in the matter era) of scalar-tensor theories with a multiplicative coupling between the scalar field and the matter Lagrangian. This class of theory parametrized by the action (\ref{eq:actiondila}) encompasses the case of the massless string dilaton considered in \cite{DamPolyGRG94} as well as the pressuron \cite{nousPRD13}. In general,  solar system constraints on the gravitation theory imply that the interesting cosmological evolutions are the one close to GR (at the exception of the pressuron that naturally satisfies the solar system tests due to a decoupling mechanism \cite{nousPRD13}). Therefore, following the procedure presented in \cite{jarvPRD12}, we have studied the cosmological evolution of these theories in the GR limit. 

First, we have considered the case where no potential is present. We have shown that the solutions depend on a critical parameter $D$ (\ref{eq:D}) that is shifted compared to the one appearing in theories with minimal coupling studied in \cite{jarvPRD12}. The solutions are therefore similar to the ones found in \cite{jarvPRD12} and depends on $D$: they can be polynomial, logarithmic or with damped oscillations. In particular, the solutions converge towards GR if $D<1$. Since the GR limit is not justified for the pressuron, we have studied the solutions in the vicinity of a constant scalar field (justified by constraints on temporal variations of the constants of Nature). We have shown that the pressuron always converge towards a constant scalar field. This is a consequence from the fact that the source term in the Klein-Gordon equation for the scalar field does not depend on the matter density but only on the pressure which vanishes in the matter era. This result has also been confirmed by a non perturbative approach.

We have also considered the case where the scalar field is self-interacting. Once again, we have used the so-called GR limit to study the cosmological evolution in the matter era. It turns out that asymptotically, the solutions do not depend on the coupling function $f(\Phi)$ between the scalar field and the matter Lagrangian. Therefore, the solutions are asymptotically exactly the same as in standard generalized Brans-Dicke whose GR limit has been studied in \cite{jarv:2010fk}. The solutions depend on a critical parameter $C$ (\ref{eq:C}) and present three behaviours, similarly to the case with no potential. In particular, a convergence towards GR appear if $A_\s W_\s<0$ (with $A_\s$ defined by (\ref{eq:defA}) and $W_\s$ by (\ref{eq:W})). Here again, the pressuron does not have to satisfy the GR limit conditions. Therefore, we have studied the evolutions of the pressuron in the vicinity of a constant scalar field. In the case of a massive pressuron, the solutions fall into two classes depending on the potential. If the potential satisfies $2V_\c=\Phi_\c V_\c'$, the behaviour of the solutions can be of three types depending on the critical parameter $B$ (\ref{eq:B}). In particular, they converge towards GR if $\frac{W'_\c}{2\omega_\c+3}\leq 0$. If the potential satisfies $V_\c=V'_\c=0$ but $V_\c''\neq 0$, then two types of solutions can appear: exponentially divergent solutions in the case where $\frac{V_\c''}{2\omega_\c+3}<0$ and damped oscillating solutions in the opposite case.

Then, we have considered the observations that are related to the late-time cosmological evolution: the constancy of the constants of Nature and the apparent acceleration of the cosmic expansion as observed by SNe Ia data. As expected, the temporal evolution of the constants of Nature is directly related to the derivative of the scalar field. Stringent constraints on the variations of the fine structure constant (and also on the gravitational constant) favour solutions converging towards a constant scalar field and exclude divergent solutions. 

Because of the non-minimal coupling between the scalar field and matter, the distance-luminosity relation is modified with respect to GR. We have derived the expression of the distance-luminosity from first principles and have shown that it explicitly depends on the coupling function $f(\Phi)$. In particular, SNe Ia data's are not sensitive to the evolution of the cosmic factor $a$ but to the effective cosmic factor $\tilde a=a\sqrt{f(\Phi)}$. Therefore, the acceleration of the effective cosmic expansion measured with SNe Ia data's is not necessarily the result of an acceleration of $a$, and it can be an effect due to the non-minimal coupling instead. 

Nevertheless, we have shown that the the conditions in order to have an acceleration of the effective cosmic factor (measured by the paremeter $\tilde q$ defined in (\ref{eq:qtilde})) are of two types: if there is no potential, the solutions have to diverge from GR; if there is a potential, the acceleration is driven by the potential (playing the role of a cosmological constant) and convergent solutions can be found. As a conclusion, the only way to produce acceleration of the effective cosmic factor while having a convergence of the scalar field in order to satisfy the constraints on the variations of the fundamental constants is to consider a potential and to keep only the converging solutions.

Even if the interesting solutions are converging towards GR, they still have small deviations from the standard $\Lambda$CDM scenario. The quantification of these deviations and the comparison with actual data is left for future work \cite{heesARXIV14}. In this context, it will be interesting to know the initial conditions at the beginning of the matter era. This requires to study the behaviour of the solutions during the radiation era. Finally, other cosmological observations (like CMB or the growth of perturbations) can also be studied to refine the current analysis.

\bibliography{PC}

\appendix

\section{String dilaton}
\label{app:dilaton}

At tree level, the dilaton is massless and couples in a universal multiplicative manner to all other fields \cite{DamPolyGRG94,*DamPolyNPB94}.  In the string representation \footnote{Aslo known as the string \textit{frame}.}, the tree level effective action directly considered in 4 dimensions writes \cite{DamPolyGRG94,*DamPolyNPB94}
\bea
S_{\rm{tree}}=\frac{1}{c} \int d^4x \sqrt{-g}&& e^{-2\Psi} \times \label{eq:treelvl}\\
&& \left(\frac{1}{2 \kappa} \left[R+ 4 \Box \Psi - 4 (\partial_\sigma \Psi)^2 \right] + \L_m \right). \nonumber
\eea
However, taking into account higher loop contributions, the coupling is expected to be modified such that \cite{DamPolyGRG94,*DamPolyNPB94,damourPRD02,*damourPRL02,damourPRD96}
\be
e^{-2\Psi} \rightarrow e^{-2\Psi} + \sum_{n=1} c_n e^{2(n-1) \Psi},
\ee 
where $n$ corresponds to the contribution of the $n^{\rm{th}}$ genus string loop. Assuming however that the coupling keeps its universality, the full loop expansion would write \cite{DamPolyGRG94}
\bea
S_{\rm{loop}}= \frac{1}{c} \int d^4x \sqrt{-g}&& B(\Psi) \times \label{eq:actionloopDP} \\
&& \left(\frac{1}{2 \kappa} \left[R+ 4 \Box \Psi - 4 (\partial_\sigma \Psi)^2 \right] + \L_m \right), \nonumber
\eea
where 
\be
B(\Psi)=e^{-2\Psi} + \sum_{n=1}^{\infty} c_n e^{2(n-1) \Psi}.
\ee
Now defining $\Phi=B(\Psi)$, assuming that $B$ is invertible such that $\Psi = A(\Phi)$, and using $\sqrt{-g} B \Box \Psi = - \sqrt{-g} B_{,\Psi} (\partial_\sigma \Psi)^2$ up to a divergence, the action (\ref{eq:actionloopDP}) can be rewritten as follows
\bea
S_{\rm{loop}}= \frac{1}{c}\int  d^4x \sqrt{-g} &&\frac{1}{2\kappa}\times   \\ 
&&\left[\Phi  R-\frac{\omega(\Phi)}{\Phi} (\partial_\sigma \Phi)^2+ 2 \kappa~\Phi~\mathcal{L}_m \right] , \nonumber
\eea
with
\be
\omega(\Phi) \equiv 4 \Phi^2 \frac{\partial A}{\partial \Phi} \left(\frac{\partial A}{\partial \Phi}+\frac{1}{\Phi} \right).
\ee

Therefore, Damour and Polyakov's dilaton \cite{DamPolyGRG94,*DamPolyNPB94} corresponds to $f(\Phi)=\Phi$ in (\ref{eq:actiondila}) --- or equivalently, $n=1$.

Note that at tree level (\ref{eq:treelvl}) however, one has $A(\Phi)=-\frac{1}{2} \ln \Phi$ and therefore $\omega=-1$. Hence,  (\ref{eq:treelvl}) is not viable as an effective action of gravitation since it cannot converge towards general relativity.

\section{Solutions of the perturbed equations}\label{app:sol}

\subsection{General case in the GR limit with no potential}\label{app:gen}
The solutions of Eqs. (\ref{eq:pertGR}) depend on a critical parameter $D$ given by (\ref{eq:D}):
\begin{itemize}
\item $D> 0$: Polynomial solution for the scalar field:
\begin{subequations}\label{eq:solDpos}
	\begin{eqnarray}
		\pm x(t)&=&\frac{1}{t}\left(M_1t^{\sqrt{D}/2} - M_2t^{-\sqrt{D}/2}\right)^2, \\
		\pm h(t)&=&\frac{2}{3t^2}\left[M_3+M_1^2\left(-a\sqrt{D}+b-\frac{c}{\sqrt{D}}\right)t^{\sqrt{D}} \right.\\
		&&\left.+M^2_2\left(a\sqrt{D}+b+\frac{c}{\sqrt{D}}\right)t^{-\sqrt{D}}+2 M_1M_2 c \ln t\right] \nonumber,
	\end{eqnarray}
\end{subequations}
where $M_{i}$ are integration constants (their expressions can be found in \cite{jarvPRD12}) and where $a,b,c$ are constants characterizing the underlying scalar-tensor theory
\begin{equation}\label{eq:abc}
	a=\frac{3+6A_\s \Phi_\s}{8A_\s\Phi_\s^2} \qquad b=\frac{3+10A_\s \Phi_\s}{8A_\s\Phi_\s^2} \qquad c=\frac{n}{\Phi_\s}.
\end{equation}
These solutions extend the ones from \cite{jarvPRD12} recovered when $n=0$. Let us mention that the behaviour depends highly on the value of $D$. 

\item $D= 0 :$ Logarithmic solution for the scalar field:
\begin{subequations}\label{eq:solD0}
	\begin{eqnarray}
		\pm x(t)&=&\frac{1}{t}\left(\tilde M_1 \ln t-\tilde M_2\right)^2,\\
		\pm h(t)&=&\frac{1}{3\Phi_*t^2}\bigg[\tilde M_3 -\frac{2}{3}\tilde M_1^2 n (\ln t)^3 \nonumber\\
		&&+\frac{1}{2}\left(\tilde M_1^2+4n\tilde M_1(2\tilde M_1+\tilde M_2)\right)(\ln t)^2\\
		&& +\left(\tilde M_1^2-\tilde M_1\tilde M_2 -2n(2\tilde M_1+\tilde M_2)^2\right)\ln t\bigg],\nonumber
	\end{eqnarray}
\end{subequations}	
where $\tilde M_{i}$ are integration constants (whose expressions can be found in \cite{jarvPRD12}). These expressions extend the ones from \cite{jarvPRD12} recovered when $n=0$. 

\item $D<0 :$ Oscillating damped solution for the scalar field:
\begin{subequations}\label{eq:solDneg}
	\begin{eqnarray}
		\pm x(t)&=&\frac{1}{t}\left[N_1\sin \left(\frac{1}{2}\sqrt{|D|}\ln t\right)\right.\nonumber \\
		&&\qquad\left. - N_2 \cos \left(\frac{1}{2}\sqrt{|D|}\ln t\right)\right],\\
		\pm h(t)&=&\frac{2}{3t^2}\Bigg\{N_3- (N_1^2+N_2^2)\frac{c}{2}\ln t +\cos (\sqrt{|D|}\ln t)\times  \nonumber\\
		&& \left[N_1N_2\left(a\sqrt{|D|}-\frac{c}{\sqrt{|D|}}\right) +\left(N_2^2-N_1^2\right)\frac{b}{2}\right]\nonumber \\
		 &+& \bigg[ (N_2^2-N_1^2)\left(\frac{a}{2}\sqrt{|D|}-\frac{c}{2\sqrt{|D|}}\right)-N_1N_2 b \bigg]\times \nonumber \\
		&& \sin (\sqrt{|D|}\ln t) \Bigg\}\, ,
	\end{eqnarray}
\end{subequations}
where $a,b,c$ are constants given by (\ref{eq:abc}) characterizing the underlying scalar-tensor theory and $N_{i}$ are integration constants whose expressions can be found in \cite{jarvPRD12}. The last expressions extend the ones from \cite{jarvPRD12}, recovered when $n=0$ (and thus $c=0$). The behaviour of these solutions is developed into details in \cite{jarvPRD12}. Basically, they approach the GR solution in the manner of damped oscillations.
\end{itemize}

\subsection{General case in the GR limit with potential}\label{app:generalPot}
The solutions of the asymptotic Eqs.~(\ref{eq:asymptotic}) are exactly the same as in \cite{jarv:2010fk}. They depend on a critical parameter $C$ given by (\ref{eq:C}):

\begin{itemize}
	\item $C > 0$ : exponential solutions:
		\begin{eqnarray}\label{eq:genPotPos}
			\pm x(t)&\infeq&e^{-\sqrt{C_1} t}\left[M_1e^{-\frac{1}{2}\sqrt{C}t}+M_2e^{\frac{1}{2}\sqrt{C}t}\right]^2.
		\end{eqnarray}
	
	\item $C=0$ : linear exponential solutions:
		\begin{eqnarray}\label{eq:genPot0}
			\pm x(t)&\infeq&e^{-\sqrt{C_1} t}\left[M_1t-M_2\right]^2.
		\end{eqnarray}

	\item $C<0$ : damped oscillating solutions
		\begin{eqnarray}\label{eq:genPotNeg}
			\pm x(t)&\infeq&e^{-\sqrt{C_1} t}\left[N_1 \sin \left(\frac{1}{2}\sqrt{|C|}t\right)\right.\\
			&&\left.-N_2\cos\left(\frac{1}{2}\sqrt{|C|}t\right)\right]^2.\nonumber
		\end{eqnarray}
\end{itemize}

\subsection{Case of the pressuron with a potential in the GR limit}\label{ap:pressuron}
The solution of the full equations (\ref{eq:limitGRPot}) with $n=1/2$ depends on the critical parameter $C$ given by (\ref{eq:C}):
\begin{itemize}
	\item $C>0$ : the solution is exponential and can be written as
	\begin{equation}
		\pm x(t)=\left(\frac{M_1e^{\sqrt{C}t/2}-M_2e^{-\sqrt{C}t/2}}{\cosh \left(K+\sqrt{\frac{3V_\s}{2\Phi_\s}}\frac{t}{2}\right)}\right)^2.
	\end{equation}

where $K$ is the integration constant appearing in (\ref{eq:Hbignonassympt}) and $M_i$ are constants of integration.

\item $C=0$: the solution is a linear exponential
\begin{equation}
	\pm x(t)=M_1\left(\frac{M_2+t}{\cosh \left(K+\sqrt{\frac{3V_\s}{2\Phi_\s}}\frac{t}{2}\right)}\right)^2.
\end{equation}

\item $C<0$: the solution is damply oscillating
\begin{equation}
	\pm x(t)=\left[ \frac{N_1\sin\left(\frac{1}{2}\sqrt{|C|}t\right)-N_2\cos\left(\frac{1}{2}\sqrt{|C|}t\right)}{\cosh \left(K+\sqrt{\frac{3V_\s}{2\Phi_\s}}\frac{t}{2}\right)} \right]^2.
\end{equation}
\end{itemize}
These solutions asymptotically converge towards the one derived in \cite{jarv:2010fk}.

\subsection{Case of the pressuron with a potential (not in the GR limit)}
\label{app:pressuronPot}.

\subsubsection{Case with $2V_\c=\Phi_\c V'_\c \neq 0$}
The asymptotically solutions of Eqs.~(\ref{eq:pressuronPotPert}) depend on the critical parameter $B$ given by (\ref{eq:B}):
\begin{itemize}
	\item $B > 0$ : exponential solutions : 
\begin{subequations}\label{eq:pressuronPotPos}
\begin{eqnarray}
x(t)&=&2\frac{M_1e^{-\frac{\sqrt{B}}{2}t} + M_2e^{\frac{\sqrt{B}}{2}t} }{\cosh \left(\frac{\sqrt{C_1}}{2}t\right)}\\
x(t)&\infeq& e^{-\frac{1}{2}\sqrt{C_1}t}\left[ M_1e^{-\frac{1}{2}\sqrt{B}t} +  M_2e^{\frac{1}{2}\sqrt{B}t}\right] \\
h(t)&\infeq&   e^{-\frac{1}{2}\sqrt{C_1}t}\Bigg[ M_1  e^{-\frac{1}{2}\sqrt{B}t} \left(\frac{\sqrt{B}}{4}+\sqrt{C_1}\frac{2+3\Phi_\c}{\Phi_\c}\right)\nonumber\\
&&  +  M_2 \ e^{\frac{1}{2}\sqrt{B}t}\left(-\frac{\sqrt{B}}{4}+\sqrt{C_1}\frac{2+3\Phi_\c}{\Phi_\c}\right)\Bigg] \\
&&\qquad+M_3 e^{-\sqrt{C_1}t}\nonumber \, , \nonumber
\end{eqnarray}
\end{subequations}

where $M_i$ are integrations constant.

\item $B=0$ or $\frac{W'_\c}{2\omega_\c+3}=-\frac{3V_\c}{8\Phi_\c}$ : linear exponential solutions :
\begin{subequations}\label{eq:pressuronPot0}
\begin{eqnarray}
x(t)&=&2\frac{M_1+M_2 t}{\cosh \left(\frac{\sqrt{C_1}}{2}t\right)}		\\
h(t)&\infeq& e^{-\sqrt{C_1}t}M_3 + e^{-\frac{1}{2}\sqrt{C_1}t}\\
&& \times \Big[\frac{2+3\Phi_\c}{12\Phi_\c}\sqrt{C_1} (M_1+M_2 t)-\frac{M_2}{2}\Big]\, ,\nonumber
\end{eqnarray}
where $M_i$ are integration constants. These solutions always converge towards a constant scalar field.
\end{subequations}

\item $B<0$ : damped oscillating solutions.
\begin{subequations}\label{eq:pressuronPotNeg}
\begin{eqnarray}
x(t)&=&2\frac{N_1\cos \left(\frac{1}{2}\sqrt{|B|}t\right)+N_2\sin \left(\frac{1}{2}\sqrt{|B|}t\right)}{\cosh \left(\frac{1}{2}\sqrt{C_1}t\right)} \\
h(t)&\infeq & e^{-\sqrt{C_1}t}N_3+ e^{-\frac{1}{2}\sqrt{C_1}t} \times\\ &&
\left[\left(\frac{2+3\Phi_\c}{12\Phi_\c}N_1\sqrt{C_1}-\frac{N_2}{4}\sqrt{B}\right)\cos\left(\frac{1}{2}\sqrt{|B|}t\right)\right.\nonumber\\
&&			\left. +\left(\frac{N_1}{4}\sqrt{B}+\frac{2+3\Phi_\c}{12\Phi_\c}N_2\sqrt{C_1}\right)\sin\left(\frac{1}{2}\sqrt{|B|}t\right)\right]\, ,\nonumber
\end{eqnarray}
\end{subequations}
where $N_i$ are constants of integration. These solutions converge towards a constant scalar field.
\end{itemize}

\subsubsection{Case with $V_\c=V_\c'=0$ and $W_\c'=-\Phi_\c V_\c''\neq 0$}

	The solution of Eqs. (\ref{eq:pressuronPotPert}) depend on a critical parameter that is simply $W_\c'=-\Phi_\c V_\c''$ 
	\begin{itemize}
		\item $\frac{W'_\c}{2\omega_c+3}>0$ or $\frac{V_\c''}{2\omega_c+3}<0$ : exponential divergent solutions:
			\begin{eqnarray}\label{eq:pressuronPot2Pos}
				x(t)&=&\frac{1}{t} \left[M_1 e^{\sqrt{\frac{W'_\c}{2\omega_c+3}}t} + M_2 e^{-\sqrt{\frac{W'_\c}{2\omega_c+3}}t} \right].
			\end{eqnarray}
			
		\item $\frac{W'_\c}{2\omega_c+3}< 0 $	or $\frac{V_\c''}{2\omega_c+3}<0$ : oscillating damped solution:
		\begin{equation}\label{eq:pressuronPot2Neg}
			x(t)=\frac{1}{t} \left[ N_1 \cos\left(\sqrt{\left|\frac{W'_\c}{2\omega_c+3}\right|}t\right)  + N_2 \sin\left(\sqrt{\left|\frac{W'_\c}{2\omega_c+3}\right|}t\right) \right].
		\end{equation}
\end{itemize}

\end{document}